\documentclass[12pt]{article}
\usepackage{graphicx}
\usepackage{amsmath}

\newcommand\pubnumber{DPF2015-149}
\newcommand\pubdate{\today}

\def\napoli{
\small on behalf of the Double Chooz collaboration ~\\
~\\
\small  Argonne National Laboratory, Lemont, Illinois 60439, USA ~\\
\small{\centering}and ~\\ 
\small Illinois Institute of Technology, Chicago, Illinois 60616, USA}

\def\Title#1{\begin{center} {\Large #1 } \end{center}}
\def\Author#1{\begin{center}{ \sc #1} \end{center}}
\def\Address#1{\begin{center}{ \it #1} \end{center}}

\newcommand\pubblock{\rightline{\begin{tabular}{l} \pubnumber\\
         \pubdate  \end{tabular}}}
\newenvironment{Abstract}{\begin{quotation}  }{\end{quotation}}
\newenvironment{Presented}{\begin{quotation} \begin{center} 
             PRESENTED AT\end{center}\bigskip 
      \begin{center}\begin{large}}{\end{large}\end{center} \end{quotation}}

\def\beq{\begin{equation}}
\def\eeq#1{\label{#1}\end{equation}}
\def\eeqn{\end{equation}}

\def\beqa{\begin{eqnarray}}
\def\eeqa#1{\label{#1}\end{eqnarray}}
\def\eeqan{\end{eqnarray}}

\let\bar=\overbar

\def\Dslash{\not{\hbox{\kern-4pt $D$}}}
\def\dslash{\not{\hbox{\kern-2pt $\del$}}}

\def\msb{{\bar{\ssstyle M \kern -1pt S}}}

\begin{document}
\begin{titlepage}
\pubblock

\vfill
\Title{Latest nH analysis in the Double Chooz experiment}
\vfill
\Author{ Guang Yang \footnote{Guang Yang: gyang9@hawk.iit.edu }} 
\Address{\napoli}
\vfill
\begin{Abstract}
Precise measurement of the neutrino mixing angle $\theta_{13}$ is the primary goal of the Double Chooz Experiment (DC), which is located
in Chooz, France. The inverse beta decay process provides a unique signature of reactor anti-neutrino interactions, giving prompt 
signals from positron annihilation and delayed signals from neutron capture by either Gadolinium (Gd) or Hydrogen (H).
This paper is dedicated to the latest nH analysis in Double Chooz. 
Typically, The Gd analysis is primary since fewer background events are involved. However, with accurate estimates of 
backgrounds and a precise reconstruction of energy, the nH analysis gives a powerful independent measurement of $\theta_{13}$.
\end{Abstract}
\vfill
\begin{Presented}
DPF 2015\\
The Meeting of the American Physical Society\\
Division of Particles and Fields\\
Ann Arbor, Michigan, August 4--8, 2015\\
\end{Presented}
\vfill
\end{titlepage}
\def\thefootnote{\fnsymbol{footnote}}
\setcounter{footnote}{0}

\section{neutrino mixing}\

Neutrinos propagate in mass states while interact in flavor states in the weak interaction. This causes the
neutrino oscillation phenomenon, which is well described by a unitary matrix introduced by Z.Maki, M.Nakagawa, S.Sakata and B.Pontecorvo (PMNS). 
The PMNS matrix is:
\begin{equation}
U_{PMNS} 
=\begin{pmatrix} c_{12}c_{13}&s_{12}c_{13}&s_{13}e^{-i\delta_{CP}} \\ 
-s_{12}c_{23}-c_{12}s_{13}s_{23}e^{i\delta_{CP}}&c_{12}c_{23}-s_{12}s_{13}s_{23}e^{i\delta_{CP}}&c_{13}s_{23} \\ 
s_{12}c_{23}-c_{12}s_{13}c_{23}e^{i\delta_{CP}}&-c_{12}s_{23}-s_{12}s_{13}c_{23}e^{i\delta_{CP}}&c_{13}c_{23} \end{pmatrix},
\label{eq:eq_1}
\end{equation}
where c$_{ij}$=cos$\theta_{ij}$ and s$_{ij}$=sin$\theta_{ij}$. $\delta_{Cp}$ is the CP-violating phase. This matrix can be broken down into
three blocks. Each of them contains one mixing angle that can be measured by certain types
of experiments. The angle $\theta_{12}$ has been precisely measured by solar neutrino experiments~\cite{SNO,BOREXINO}. The angle $\theta_{23}$ is being measured by 
beam and atmospheric neutrino experiments~\cite{superK,MINOS} and $\theta_{13}$ is being measured by reactor antineutrino experiments like Double Chooz~\cite{DC}, Daya Bay~\cite{DB} and RENO~\cite{RENO}.
The major remaining puzzles of neutrino oscillation are the sign of $\Delta m_{31}^{2}$, the CP violation phase, the $\theta_{23}$ octant and the existence of Majorana neutrinos.

\section{Double Chooz experiment}\

Double Chooz is located in Chooz, France and it started the data taking in 2012. Two detectors
are designed to take data but for the nH analysis presented here, only far detector data are used. 
The far detector has an average distance 1050 m from the reactor cores. The total live time in
the data sample is 472.72 days. The signal Double Chooz uses is inverse beta decay (IBD). When an electron
antineutrino from the reactors enters the detector, it interacts with a proton, then a positron and a 
neutron are created. The positron can be observed as a prompt signal stemming from its ionization and annihilation with an electron, 
while the neutron can be observed as a delayed signal if it is captured by hydrogen or gadolinium.
The survival probability of electron antineutrino travelling around 1 km can be approximated as:
\begin{equation}
P(\overline{\nu}_{e}\rightarrow\overline{\nu}_{e})=1-sin^{2}(2\theta_{13})sin^{2}(\frac{\Delta m^{2}_{31} L}{4E}),
\label{eq:eq_2}
\end{equation}
where L is the baseline and E$_{\nu}$ is the neutrino energy.\

The Double Chooz far detector consists of four concentric cylindrical vessels. The outermost volume is called the inner veto (IV).
It is filled with liquid scintillator and equipped with 78 8-inch photomultiplier tubes (PMTs) so that it operates
as a muon veto and a shield. The second outermost volume is called the buffer and it is filled with mineral oil.
The inner detector consists of the neutrino target (NT) and gamma catcher (GC). Both of them are filled with liquid scintillator
but NT is also doped with gadolinium, which has a large cross section to capture neutrons.\

In the latest nH analysis, Double Chooz improves several important things over previous papers to provide more precise measurement, including
energy reconstruction, background reduction, detection systematics evaluation and the final fit strategy. These will be
briefly introduced one by one in the following sections.\

\section{Energy reconstruction}\

\begin{figure}
\centering
\includegraphics[angle=0,width=13cm,height=5cm]{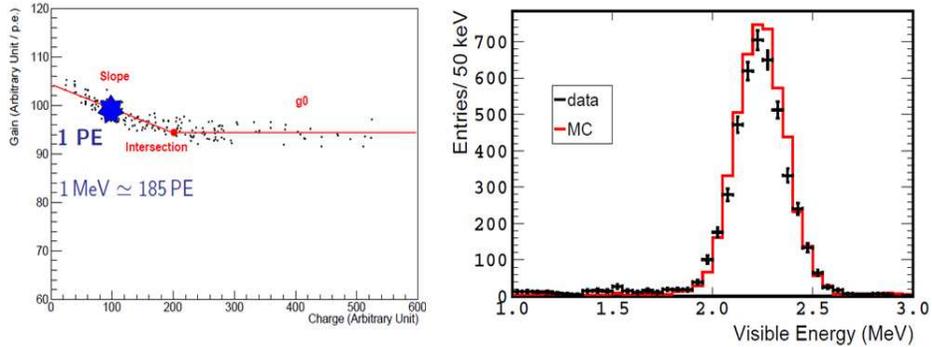}
\caption{{Left: The gain calibration. Right: The PE/MeV calibration using
the hydrogen capture peak from the radioactive source $^{252}$Cf.}}
\label{fig:fig_energy}
\end{figure}

The energy reconstruction for DC nH analysis can be described as:
\begin{equation}
E_{vis} = N_{pe} \times f_{u}(\rho,z) \times f_{PE/MeV} \times f_{s}^{data}(E^{0}_{vis},t) \times f^{MC}_{nl}.
\label{eq:eq_3}
\end{equation}
The first factor is the gain calibration, which contains the information of transforming the total charge
to the number of photoelectrons. The left panel on fig.~\ref{fig:fig_energy} shows the gain calibration.
The second factor is the spatial uniformity calibration. By using the
spallation neutron captures in hydrogen, we get uniformity maps for data and MC separately. The third
factor is the calibration that transforms the number of photoelectrons to visible energy. We use
hydrogen captures from radioactive source $^{252}$Cf for this. The right panel on fig.~\ref{fig:fig_energy}
shows this calibration. The fourth factor is the time stability calibration, which is implemented by using
natural radioactive sources. The last factor accounts for the light and charge nonlinearity, arising from 
the remaining discrepancy between data and MC. It 
is only applied to MC, to make it to be consistent with data. 
\

\section{Background reduction and detection systematics determination}\

\begin{figure}
\centering
\includegraphics[angle=0,width=13cm,height=5cm]{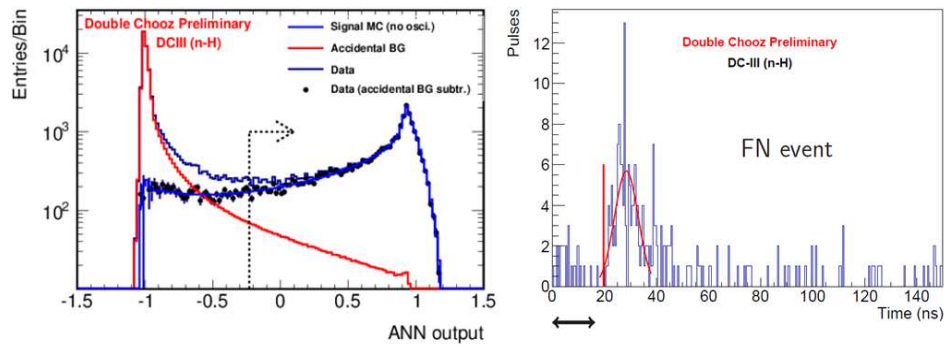}
\caption{{Left: ANN output values for: Black histogram: on-time data, Red: Off-time data, Black points:
on-time minus off-time data, Blue: signal MC. Right: The pulse shape for a fast neutron event.}}
\label{fig:fig_background}
\end{figure}

Basically, there are three kinds of backgrounds considered in Double Chooz. They are accidental background,
fast neutron/stopping muon and $^{9}$Li/$^{8}$He. Artificial neural network (ANN) is a powerful tool to 
reduce the accidental background, which is the main contamination in the nH analysis. We use the time difference between the
prompt and delayed signals, distance difference and the delayed energy as inputs to the MLP (Multi-layer Perceptron)
network. ANN is trained by using background samples from data and signal samples from MC. The left panel on fig.~\ref{fig:fig_background}
shows the ANN output values for signal MC, accidental background and data.\

The fast neutron background can be eliminated by a new technique named MPS (multiple pulse shape). Proton recoils from the fast neutron events 
can cause a small shift on the pulse shape. Removing those events will reduce the fast neutron background effectively. The right panel on 
fig.~\ref{fig:fig_background} shows the pulse shape of a fast neutron event as an example.\
\

The detection systematics is carefully computed by taking three factors into account. First, the hydrogen fraction
accounts for the fraction of events captured by hydrogen, rather than captured by gadolinium. This is calculated by
using $^{252}$Cf and IBD events happening at the target center and at positions in the gamma catcher which are far away from the target.
Those two cases provide two numbers for the target region and gamma catcher region separately.
Second, the spill in/out uncertainty is estimated by calculating the difference between two simulation tools, TRIPOLI-4 and GEANT4.
Last, the proton number uncertainty is also considered.

\section{Final fit}\

\begin{figure}
\centering
\includegraphics[angle=0,width=13cm,height=5cm]{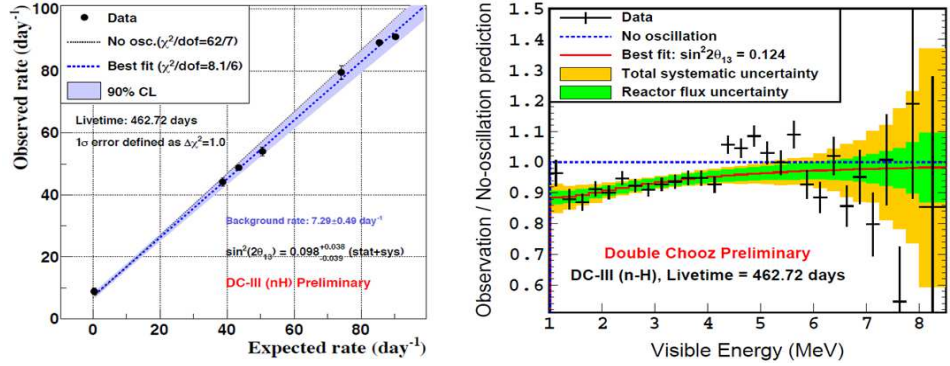}
\caption{{Left: Observed rates vs. expected rates for the 7 reactor rate modes. Best fit line is shown as a dashed line. Right: Observation and non-oscillation prediction ratio along the visible
energy. The red histogram shows the best fit curve.}}
\label{fig:fig_FF}
\end{figure}

The next step is to proceed the final fit. We have two final fit methods. The first one is the RRM (reactor rate module) fit.
The reactors which generate the electron antineutrinos have six reactor rate modes plus one reactor-off mode during the data taking period. By using these modes,
one can perform a fit relying on the rate-only information.  The second one is the rate+shape fit. This is a traditional fit with 
covariance matrices and pull terms. The pulls include background constraints, $\Delta m^{2}_{31}$, energy scale parameters and the residual number
of neutrinos on the reactor-off period. The best RRM fit gives $sin^{2}(2\theta_{13})$=0.098$^{+0.038}_{-0.039}$ with a background
rate constraint while the best rate+shape fit gives $sin^{2}(2\theta_{13})$=0.124$^{+0.030}_{-0.039}$. These results
are consistent with the results we measured before~\cite{DC}. The left panel on fig.~\ref{fig:fig_FF} shows the observed rate vs. the expected rate from the RRM fit and the right panel
shows the ratio of observation to the no-oscillation prediction vs. the visible energy. A distortion
around 5 MeV is observed and has similar feature as observed before. This data excess contributes significantly to the $\chi^{2}$ value.
\

\section{Conclusion}\

Double Chooz performed an independent hydrogen capture analysis and it gives consistent results to the gadolinium capture analysis.
With 472.72 days data, the RRM fit and rate+shape fit are performed and they give consistent results, $sin^{2}(2\theta_{13})$=\\0.098$^{+0.038}_{-0.039}$.
As of the summer of 2015, the near detector has started data taking and we are preparing for a six-month data set. With both detectors,
the sensitivity to $\theta_{13}$ is expected to improve significantly due to the enormous cancellation of systematics and the increase of statistics.
With this contribution from Double Chooz, $\theta_{13}$ is the most precisely measured parameter in the neutrino oscillation currently.

\end{document}